\input{aipcheck.tex}
\documentclass{aipproc}
\layoutstyle{8s} 
\begin{document}
\title [CMB Polarization: Scientific Case and Data Analysis Issues]
{CMB Polarization: Scientific Case and Data Analysis Issues}
\author{A. Balbi$^*$}{ address={INFN, Sezione di Roma II},
  address={Dipartimento di Fisica, Universit\`a Tor Vergata, Roma
    I-00133, Italy}, }
\author{P. Cabella}{ address={Dipartimento di Fisica, Universit\`a Tor
    Vergata, Roma I-00133, Italy}, }
\author{G.  de Gasperis}{ address={Dipartimento di Fisica,
    Universit\`a Tor Vergata, Roma I-00133, Italy}, }
\author{P.  Natoli$^*$}{ address={INFN, Sezione di Roma II},
  address={Dipartimento di Fisica, Universit\`a Tor Vergata, Roma
    I-00133, Italy}, }
\author{N. Vittorio$^*$}{ address={INFN, Sezione di Roma II},
  address={Dipartimento di Fisica, Universit\`a Tor Vergata, Roma
    I-00133, Italy}, }
\begin{abstract}
  We review the science case for studying CMB polarization.  We then
  discuss the main issues related to the analysis of forth-coming
  polarized CMB data, such as those expected from balloon-borne (e.g.
  BOOMERanG) and satellite (e.g. Planck) experiments.
\end{abstract}
\date{\today}
\maketitle
%
%======
\section{Introduction}
%======
%
Strong theoretical arguments suggest the presence of fluctuations in
the polarized component of the cosmic microwave background (CMB) at a
level of 5-10\% of the temperature anisotropy. A wealth of scientific
information is expected to be encoded in this polarized signal.
However, while the existence of anisotropies in the temperature of the
Cosmic Microwave Background (CMB) has now been firmly established by
several experiments \cite{boom, max, dasi}, only upper limits are
currently available for fluctuations in the polarization of the CMB
radiation.

The prospect of detecting CMB polarization anisotropy at small angular
scales is now more promising than in the past. In the next few years,
a number of experiments (e.g. BOOMERanG, Planck) will have the right
sensitivity, as well as the necessary control on systematic effects, to
make the measurement of polarization an achievable goal.

In this contribution we will first quickly review the major features
of CMB polarization, then we will address some of the issues that will
have to be faced in order to analyze the data collected by the
forthcoming experiments.
%
%======
\section{CMB Polarization: Theoretical Framework}
%======
%
There are at least three features of CMB polarization that make it an
appealing target for observation.  First, CMB polarization
anisotropies are generated at last scattering, so they are not
affected by effects taking place after recombination, as opposed to
temperature anisotropies.  Second, distinctive polarization patterns
are produced by different kind of density perturbations (e.g. scalar,
vector, tensor). Finally polarization provides information
complementary to temperature, helping in clarifying issues such as
cosmological parameter degeneracies in the temperature power spectrum.

The rest of this chapter covers basic theoretical aspects of CMB
polarization. Excellent reviews on the subject are \cite{koso,primer}.
%
%------
\subsection{Formalism}
%------
%
A useful way to characterize the polarization properties of the CMB is
to use the Stokes parameters formalism \cite{chandra}. For a nearly
monochromatic plane electromagnetic wave propagating in the $z$
direction,
\begin{equation}
  E_x=a_x(t)\cos\left[\omega_0 t - \theta_x(t)\right], \ \ \ \ \ \
  E_y=a_y(t)\cos\left[\omega_0 t - \theta_y(t)\right],
\end{equation}
the Stokes parameters are defined by:
\begin{equation}
  I\equiv \langle a_x^2\rangle + \langle a_y^2\rangle, \ \ \ \ \ \ \ \
  Q\equiv \langle a_x^2\rangle - \langle a_y^2\rangle, \ \ \ \ \ \ \ \
  U\equiv \langle 2a_xa_y\cos(\theta_x -\theta_y)\rangle, \ \ \ \ \ \ \ \
  V\equiv \langle 2a_xa_y\sin(\theta_x -\theta_y)\rangle,
\end{equation}
where the brackets $\langle \rangle$ represent time averages. The
parameter $I$ is simply the average intensity of the radiation. The
polarization properties are described by the remaining parameters: $Q$
and $U$ describe linear polarization, while $V$ describes circular
polarization.  Unpolarized radiation (or natural light) is
characterized by having $Q=U=V=0$.  CMB polarization is produced
through Thomson scattering (see below) which, by symmetry, cannot
generate circular polarization. Then, $V=0$ always for CMB
polarization.  The Stokes parameters $Q$ and $U$ are not scalar
quantities. If we rotate the reference frame of an angle $\phi$ around
the direction of observation, $Q$ and $U$ transform as:
\begin{equation}
  Q'=Q\cos(2\phi) + U\sin(2\phi), \ \ \ \ \ \
  U'=-Q\sin(2\phi) + U\cos(2\phi).
\end{equation}
We can define a {\em polarization vector\/} ${\bf P}$ having:
\begin{equation}
  \left\vert {\bf P}\right\vert = \left(Q^2+U^2\right)^{1/2}, \ \ \ \ \ 
  \alpha={1\over 2}\tan^{-1}\left({U\over Q}\right).
\end{equation}
Although ${\bf P}$ is a good way to visualize polarization, it is not
properly a vector, since it remains identical after a rotation of
$\pi$\/ around $z$, thus defining an orientation but not a direction.
Mathematically, $Q$ and $U$ can be thought as the components of the
second-rank symmetric trace-free tensor:
\begin{equation}
  {\bf P}_{ab}={1\over 2} \left( \begin{array}{cc}
   Q & -U \sin\theta \\
   \noalign{\vskip6pt}
   - U\sin\theta & -Q\sin^2\theta \\
   \end{array} \right),
\end{equation}
where the trigonometric functions come from having adopted a spherical
coordinate system.
%
%------
\subsection{Physical Mechanisms}
%------
%
The CMB photons interact before recombination with the free electrons
of the primeval plasma through Thomson scattering. The dependence of
the Thomson scattering cross-section on polarization is given by:
\begin{equation}
{d\sigma\over d\Omega}={3\sigma_T\over 8\pi} \left\vert
   \hat\epsilon\cdot\hat\epsilon'\right\vert^2,
\end{equation}
where $\hat\epsilon$ and $\hat\epsilon'$ are incident and scattered
polarization directions.
After scattering, initially unpolarized light has: 
\begin{equation}
  I={3\sigma_T\over 16\pi}I'\left(1+\cos^2\theta\right), \ \ \ \ \ \
  Q={3\sigma_T\over 16\pi}I'\sin^2\theta, \ \ \ \ \ \
  U=0.
\end{equation}
Integrating over all incoming directions gives:
\begin{equation}
  I={3\sigma_T\over 16\pi}
  \left[{8\over 3}\sqrt{\pi}\, a_{00} + {4\over 3}\sqrt{\pi\over 5} a_{20}
  \right], \ \ \ \ \ \
  Q-iU={3\sigma_T\over 4\pi} \sqrt{2\pi\over 15} a_{22}.
\end{equation}
Then, polarization is only generated when a quadrupolar anisotropy in
the incident light at last scattering is present. This has two
important consequences. Because it is generated by a causal process,
CMB polarization peaks at scales smaller than the horizon at last
scattering. Moreover, the degree of polarization depends on the
thickness of last scattering surface. As a result, the polarized
signal for standard models at angular scales of tens of arcminutes is
about $10\%$ of the total intensity (even less at larger scales).
Typically, this means a polarized signal of a few $\mu$K.
%
%------
\subsection{Statistics}
%------
%
Since polarization is not a scalar, it cannot be expanded over the sky
using spherical harmonics, as it is done with temperature. We can
however expand the polarization tensor using the {\em tensor spherical
  harmonics\/} basis \cite{varsha}, as:
\begin{equation}
  {\bf P}_{ab} =  T_0\sum_{l=2}^\infty\sum_{m=-l}^l \left[
    a_{(lm)}^{\rm E} Y_{(lm)ab}^{\rm E} + a_{(lm)}^{\rm B}
    Y_{(lm)ab}^{\rm B}  \right]
\end{equation}
where the $E$ and $B$ labels refer to the scalar and pseudo-scalar
components of the polarization tensor.  The statistical properties of
the CMB anisotropies polarization are then characterized by six power
spectra: $C_l^T$ for the temperature, $C_l^E$ for the E-type
polarization, $C_l^B$ for the B-type polarization, $C_l^{TE}$,
$C_l^{TB}$, $C_l^{EB}$ for the cross correlations.
For the CMB, $C_l^{TB}=C_l^{EB}=0$. Furthermore, since $B$ relates to
the component of the polarization field which possesses a handedness,
one has $C_l^{B}=0$ for scalar density perturbations. The detection of
a non zero $B$ component would point to the existence of a tensor
contribution to density perturbations.

The relation relating ($E$,$B$) to ($Q$,$U$) has a non-local nature.
In the limit of small angles it can be written as:
\begin{equation}
  E(\bf{\theta})=-\int d^2{\bf \theta}^{\prime}\
  \omega(\tilde \theta)\ Q_r({\bf \theta}^{\prime}), \ \ \ \ \ \ 
  B(\bf{\theta})=-\int d^2{\bf \theta}^{\prime}\
  \omega(\tilde \theta)\ U_r({\bf \theta}^{\prime}),
\end{equation}
where the 2D angle $\bf{\theta}$ defines a direction of observation in
the coordinate system perpendicular to $z$,
\begin{equation}
  Q_r({\bf \theta})=Q({\bf \theta}^{\prime})
  \cos(2\tilde\phi) - U({\bf \theta}^{\prime})
  \sin(2\tilde\phi), \ \ \ \ \ \ \ 
  U_r({\bf \theta})=U({\bf \theta}^{\prime})
  \cos(2\tilde\phi) + Q({\bf \theta}^{\prime})
  \sin(2\tilde\phi).
\end{equation}
and $\omega(\tilde \theta)$ is a generic window function.
%
%------
\subsection{Theoretical Predictions}
%------
%
\begin{figure}[!t]
  \includegraphics[height=.3\textheight]{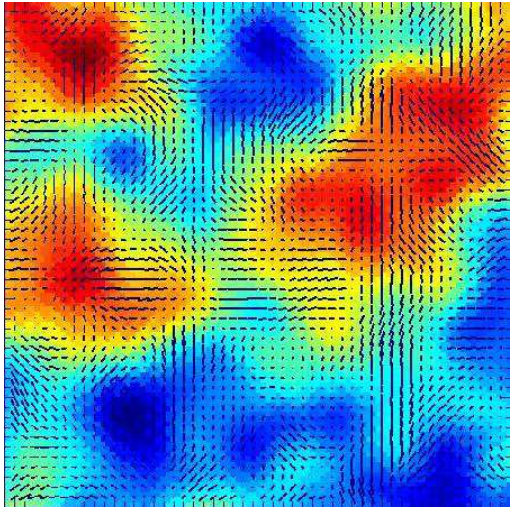}
  \caption{A simulated CMB temperature map and the corresponding polarized 
    component represented by the polarization vector $\vert{\bf
      P}\vert$, for a standard cosmological model where only scalar
    density perturbations are present.  The field is $6^\circ\times
    6^\circ$, the resolution is $10^\prime$ FWHM.}\label{polamap}
\end{figure}
The theoretical study of the CMB, for what concerns both its polarized
and unpolarized components, is in a fully mature stage. We can produce
high-precision predictions of the expected statistical CMB pattern for
any given cosmological model.  Figure \ref{polamap} shows a simulated
map of the CMB temperature anisotropy, and the corresponding
polarization field, represented by the polarization vector $\vert{\bf
  P}\vert$. This kind of simulation can be very helpful in
investigating optimal observational strategy for future experiments.
In particular, now that high-resolution CMB temperature maps are
available for certain areas of the sky, one can use this information
to predict the statistical properties of the expected polarized signal
in those regions, and tailor the polarization observations to enhance
the likelihood of a detection.
\begin{figure}[!t]
  \includegraphics[height=.3\textheight]{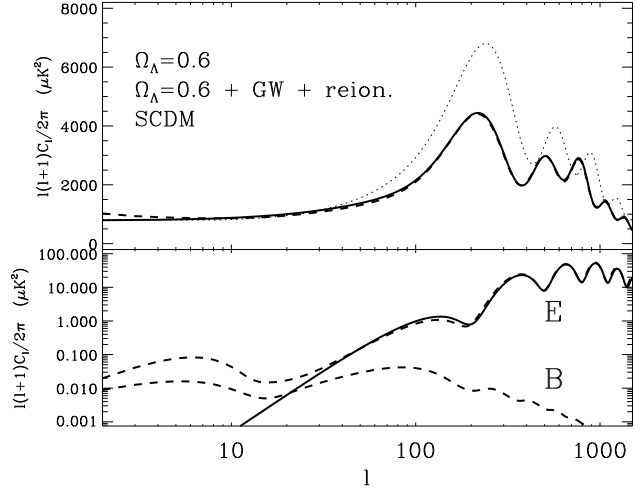}
  \caption{{\em Top panel\/} -- Temperature power spectra for a standard 
    CDM model (solid line), and the same CDM model with a fraction of
    the critical density coming from a cosmological constant (dotted
    line). If we add to the latter a contribution from tensor
    perturbations (gravitational waves background) and reionization
    (dashed line) we can make it indistinguishable from the standard
    CDM model. {\em Bottom panel\/} -- The same models and their
    polarization power spectra. The CDM model can be identified by its
    polarized signal, because it does not generate a B-type
    component}\label{polaps}
\end{figure}
Figure \ref{polaps} shows an example of how a polarization measurement
could complement the information from temperature. Two models that
would be undistinguishable from their temperature power spectra
(because of the degeneracy between the effect of reionization and of
tensor modes) can be discerned by their signature in the polarization
power spectrum: only the model having a tensor contribution produces
B-type polarization. Furthermore, reionization produces a bump at low
$l$'s in the polarization spectrum.
%
%======
\section{CMB Polarization: Data Analysis}
%======
%
In order to extract the cosmological information encoded in the CMB
polarization, one has to face the challenge of a complicated data
analysis stage after the observations are performed. CMB data analysis
was successfully performed for recent CMB temperature experiments.
While analyzing polarized data is in principle not different from
unpolarized data, further complications have to be addressed. In the
following we show how the problem of producing polarized maps from
time-ordered observations can be addressed using the same kind of
algorithms and formalism developed for the temperature case \cite{natoli}.
%
%------
\subsection{Map-making}
\begin{figure}[!ht]
    \resizebox{.95\textwidth}{!} {\includegraphics{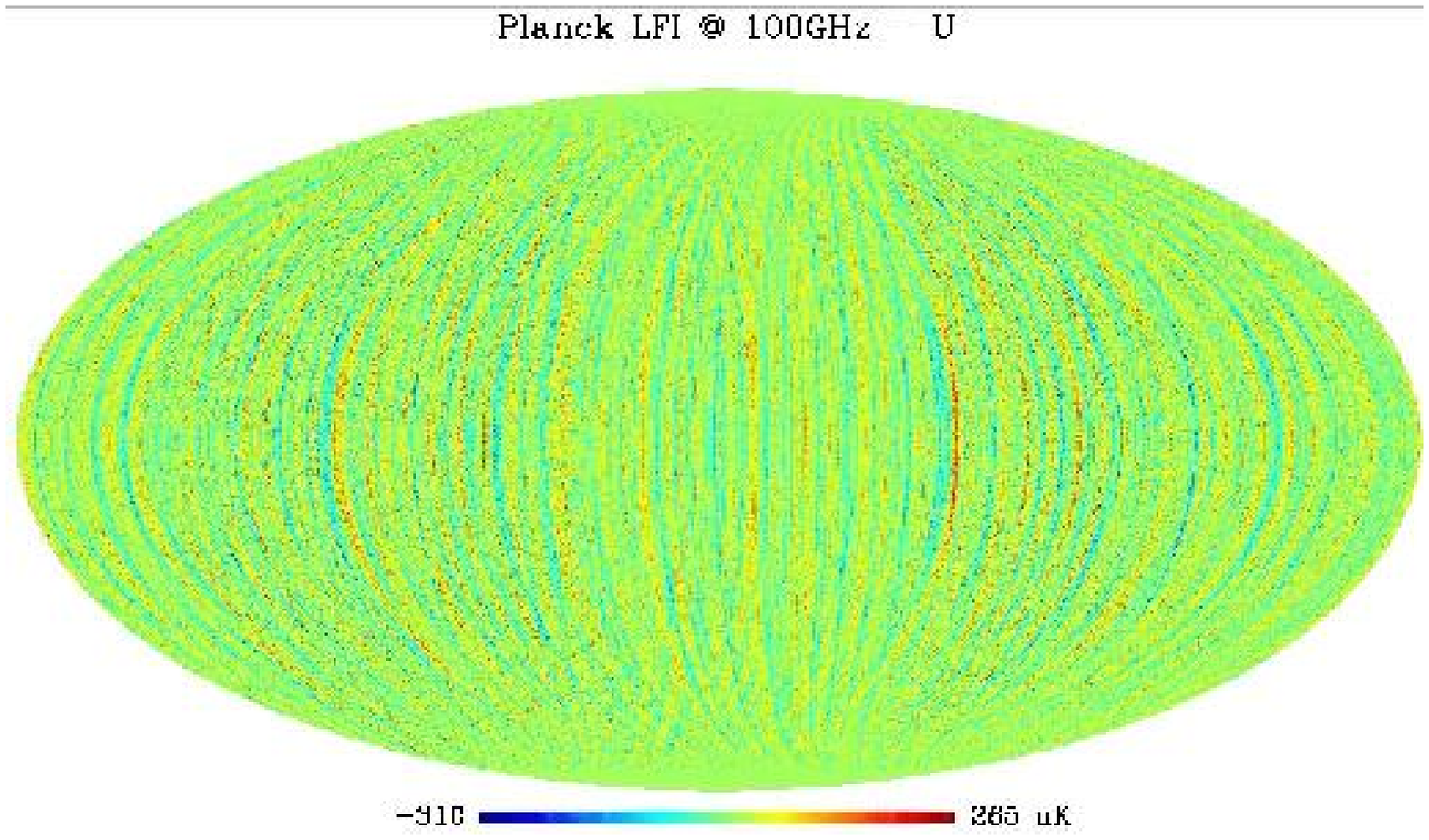} 
    \includegraphics{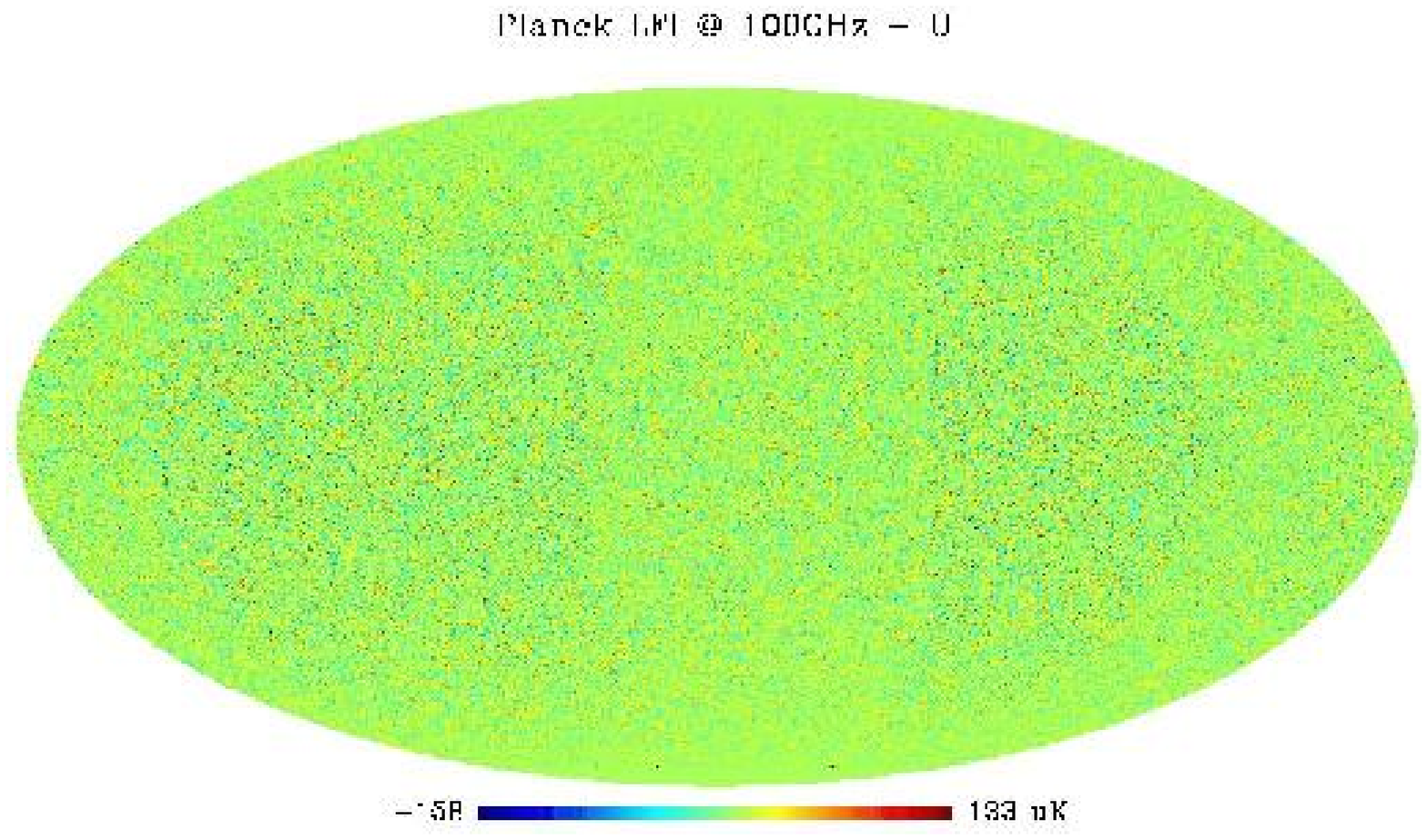}}
  \caption{Simulated maps for the 100 GHz channels (32 radiometers) 
    of Planck/LFI. Shown from top to bottom are the I, Q and U
    components, obtained by a naive coadding of observations in each
    pixel (left) and using the map-making procedure described in the
    text (right).}\label{mapmak}
\end{figure}

%------
%
We can write the total signal measured from a generic noiseless
polarimeter as:
\begin{equation}
  \mathcal M = \frac{1}{2}\left( I + Q \cos 2\phi + U \sin2\phi \right).
\end{equation}
\begin{figure}[!h]
  \includegraphics[height=.15\textheight]{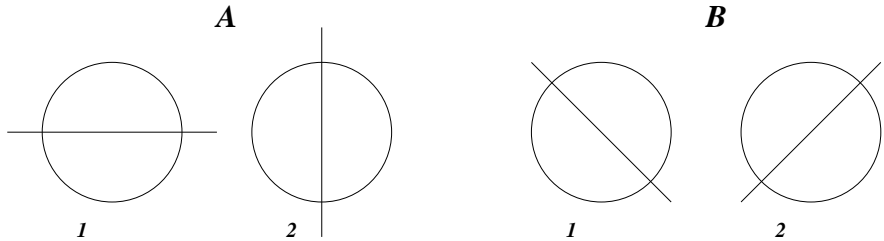}
  \caption{Experimental set-up for two radiometers, each measuring 
    two polarization states at $90^\circ$ orientation. Furthermore,
    the radiometer B polarimeters are at $45^\circ$ with respect to
    radiometer A}\label{radiometers}
\end{figure}
Consider the experimental set-up shown in Figure \ref{radiometers},
where the four polarimeters are assumed to observe the same point on
the sky. Then, one can write:
\begin{eqnarray}
  Q &=& \left( \mathcal M_{A1} - \mathcal M_{A2}\right)\cos 2\phi - 
  \left( \mathcal M_{B1} - \mathcal M_{B2}\right)\sin 2\phi \\
  U &=& \left( \mathcal M_{A1} - \mathcal M_{A2}\right)\sin 2\phi + 
  \left( \mathcal M_{B1} - \mathcal M_{B2}\right)\cos 2\phi.
\end{eqnarray}
More generally, the time-ordered data stream for a noisy polarimeter
$(i)$ is:
\begin{equation}
  d_t^{\left( i\right)} = \frac{1}{2} A_{tp}^{\left( i\right)}
  \left[ I + Q_p \cos 2\phi_t^{\left( i\right)} + 
    U_p \sin 2\phi_t^{\left( i\right)} \right] + n_t^{\left( i\right)}.
\end{equation}
For the previous set-up: 
\begin{eqnarray}
  \widetilde{d_t}^A 
  & = & A_{tp}^{A}\left[ Q_p\cos 2\psi_t + U_p\sin 2\psi_t \right] + 
  \widetilde{n_t}^A \\
  \widetilde{d_t}^B
  & = & A_{tp}^{B} \left[ -Q_p\sin 2\psi_t + U_p\cos 2\psi_t \right] + 
  \widetilde{n_t}^B, \\
\end{eqnarray}
where:
\begin{equation}
\widetilde{d_t}^A \equiv d_t^{A1} - d_t^{A2},\ \ \ \ \ \
\widetilde{d_t}^B \equiv d_t^{B1} - d_t^{B2},\ \ \ \ \ \
\widetilde{n_t}^A \equiv  n_t^{A1} - n_t^{A2},\ \ \ \ \ \
\widetilde{n_t}^B \equiv  n_t^{B1} - n_t^{B2}.
\end{equation}
We can recast everything in a matrix formalism:
\begin{equation}
  \mathbf D_t = \mathbf A_{tp} \mathbf S_p + \mathbf n,
\end{equation}
where:
\begin{equation}
  \mathbf D_t \equiv
  \left(
    \begin{array}{c}
      \widetilde{d_t}^A \\
      \widetilde{d_t}^B
    \end{array}
  \right),\ \ \ \ \ \
  \mathbf A_{tp} \equiv
  \left(
    \begin{array}{cc}
      \cos 2 \psi_t A_{tp}^A & \sin 2 \psi_t A_{tp}^A \\
      & \\
      -\sin 2 \psi_t A_{tp}^B & \cos 2 \psi_t A_{tp}^B
    \end{array}
  \right),\ \ \ \ \ \
  \mathbf S_p \equiv 
  \left(
    \begin{array}{c}
      Q_p \\
      U_p
    \end{array}
  \right),\ \ \ \ \ \
  \mathbf n_t \equiv
  \left(
    \begin{array}{c}
      n_t^A \\
      n_t^B
    \end{array}
  \right).
\end{equation}
We then obtain the standard map-making solution:
\begin{equation}
  \mathbf {\widetilde S_p} = \left( \mathbf A^t \mathbf{N}^{-1} \mathbf A\right)
  \mathbf A^t \mathbf{N}^{-1} \mathbf{D},
\end{equation}
where:
\begin{equation}
  \mathbf N \equiv \left\langle \mathbf n_t \mathbf n_{t^\prime}\right\rangle=
  \left(
    \begin{array}{cc}
      \left\langle  n_t^A  n_{t^\prime}^A \right\rangle & 
      \left\langle  n_t^A  n_{t^\prime}^B \right\rangle\\
      & \\
      \left\langle  n_t^B  n_{t^\prime}^A \right\rangle &
      \left\langle  n_t^B  n_{t^\prime}^B \right\rangle
    \end{array}
  \right),
\end{equation}
which can be further simplified if: 
\begin{equation}
  \left\langle n_t^A n_{t^\prime}^B
  \right\rangle\ = \left\langle n_t^B n_{t^\prime}^A \right\rangle =
  \mathbf 0.
\end{equation}
An example of the application of this procedure is shown in Figure
\ref{mapmak}, for the 100 GHz channel of Planck Low Frequency
Instrument.
%
%======
\section{Conclusions}
%======
%
Temperature anisotropy measurements have just started to have the
accuracy required for high precision cosmology. Polarization has
enormous scientific potential, but is still a big challenge, both
experimentally (low signal, fine-scale structure, systematics, etc.)
and for data analysis (which must be both accurate and efficient).
The next few years will most likely bring us definitive
high-resolution high-sensitivity maps of the CMB temperature
anisotropy by satellites (MAP, Planck). The new frontier of
cosmological exploration will then shift towards observations of CMB
polarization, which will certainly provide us with new insights about
the physics of the early universe.

\begin{figure}[!ht]
    \resizebox{.95\textwidth}{!}  {\includegraphics{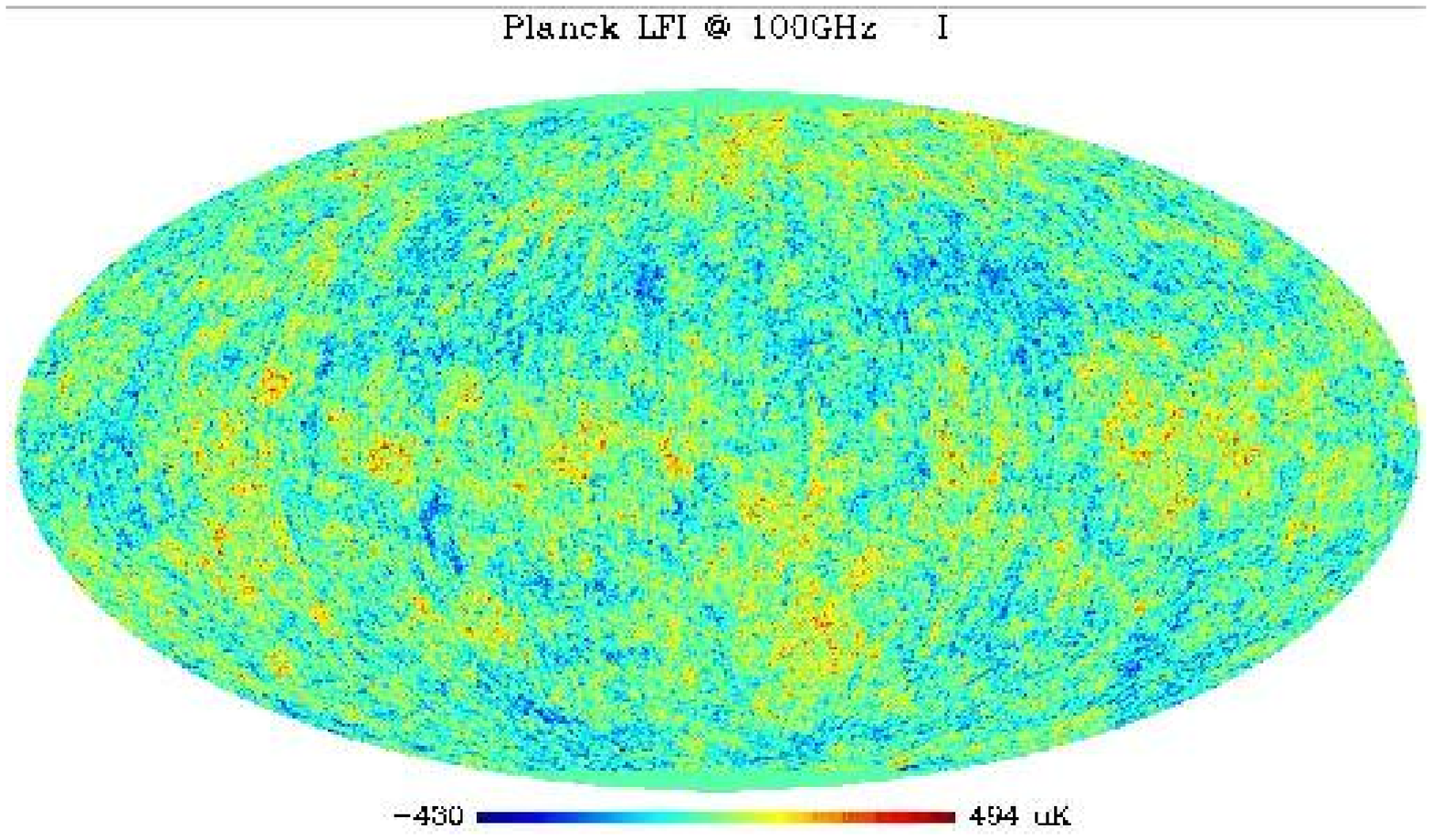}
    \includegraphics{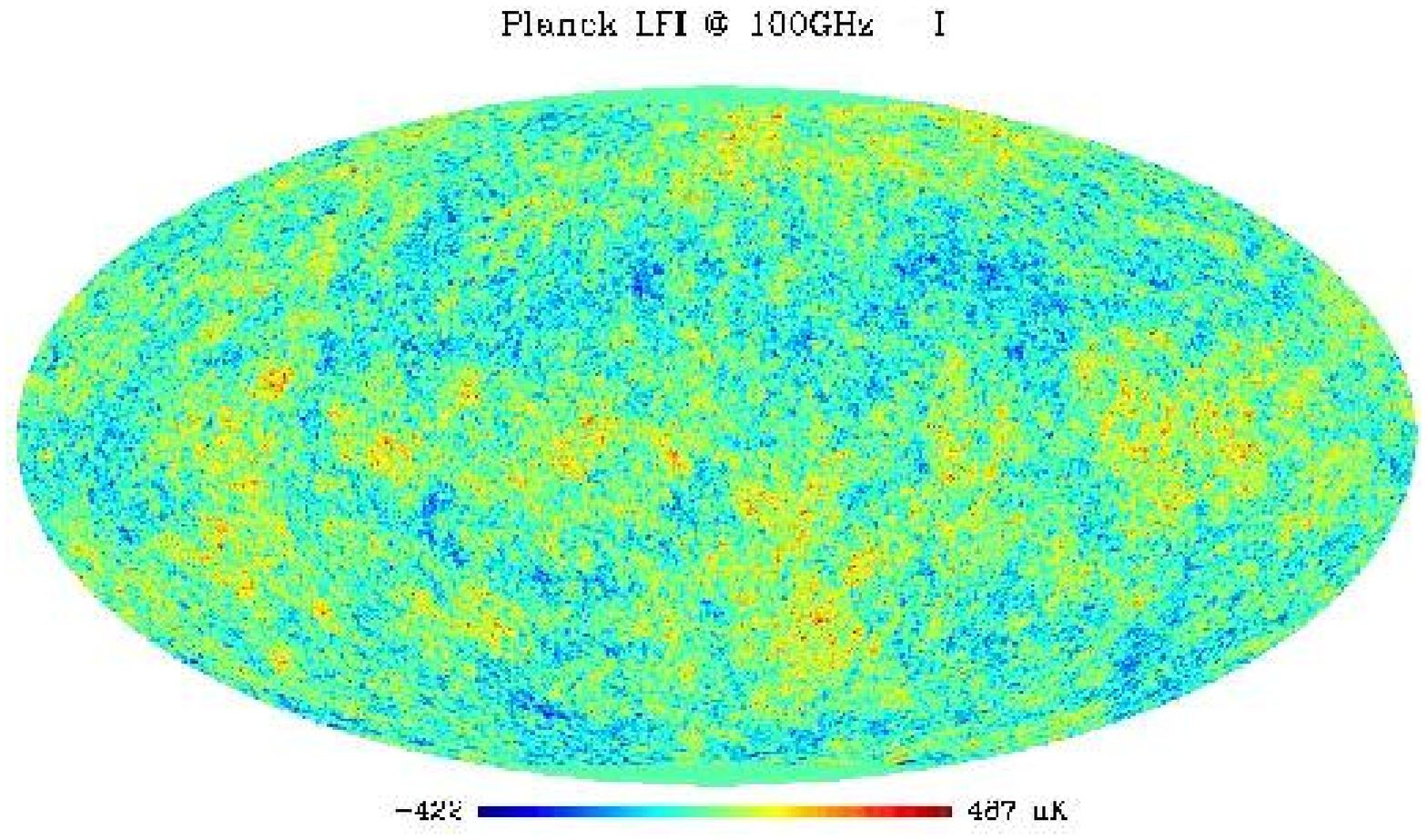}}
\end{figure}    
\begin{figure}[!ht]
    \resizebox{.95\textwidth}{!} {\includegraphics{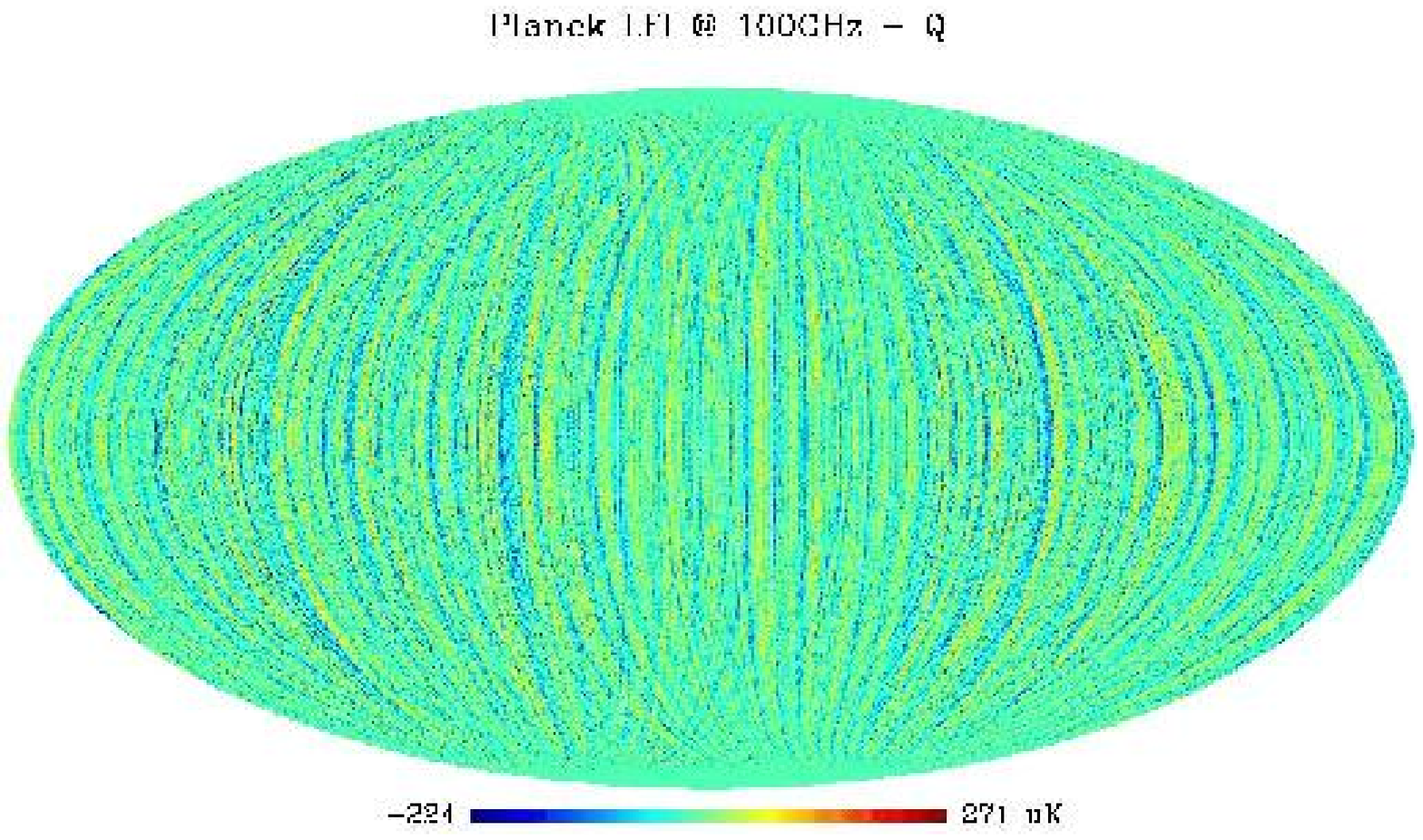} 
    \includegraphics{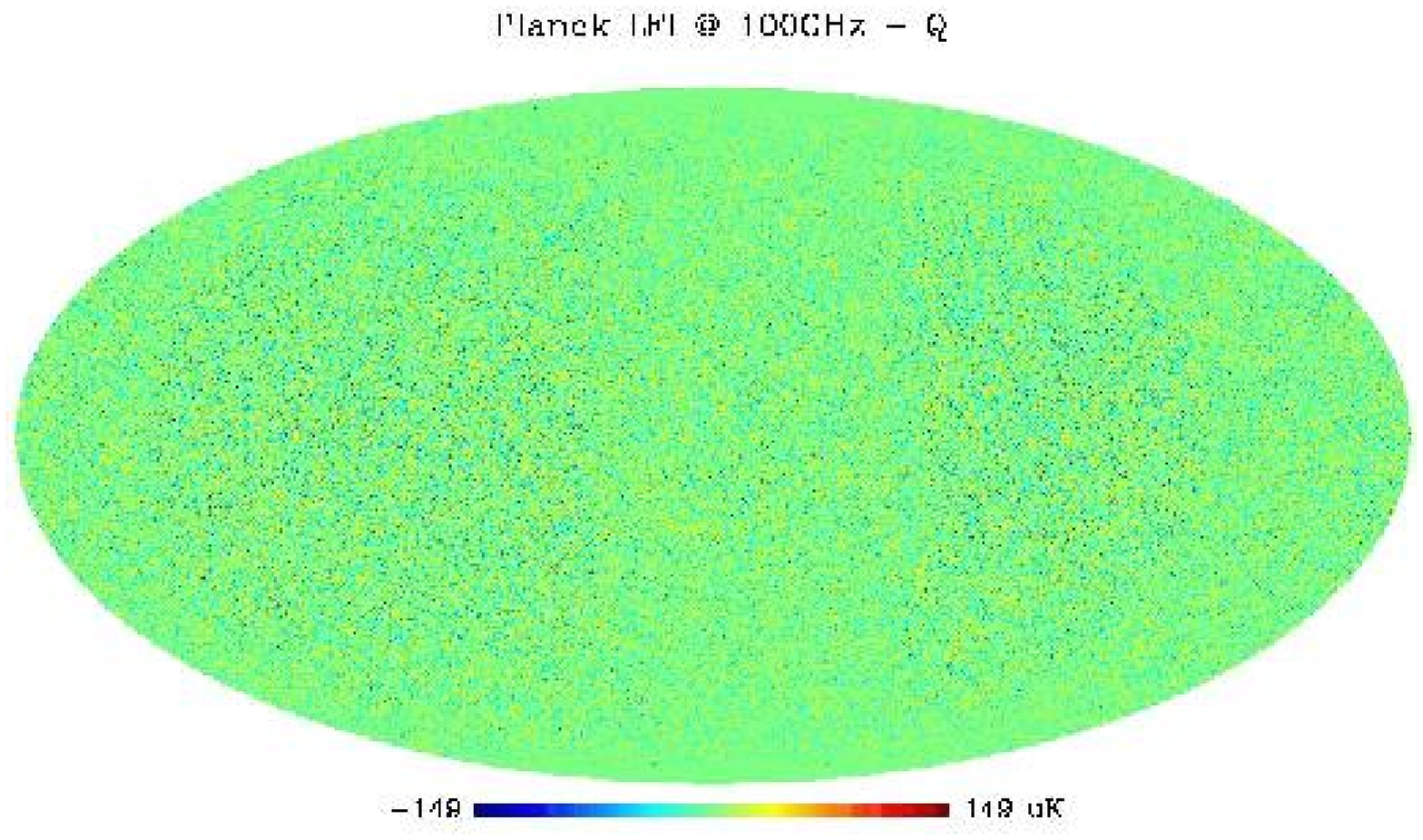}}
\end{figure}    
%
%======
\begin{theacknowledgments}
  It is a pleasure to thank the organizers of this interesting
  workshop for the invitation and for the stimulating environment. We
  acknowledge use of HEALPix ({\tt
    http://www.eso.org/science/healpix/}) and CMBFAST.
\end{theacknowledgments}
%======
%
%======

%======
%

\end{document}